# Quantum model for electro-optical amplitude modulation

**José Capmany[1,*] and Carlos R. Fernández-Pousa[2]**

[1]*ITEAM Research Institute, Universidad Politécnica de Valencia, C/ Camino de Vera s/n, 46022 Valencia, Spain*
[2]*Signal Theory and Communications, Department of Physics and Computer Science, Universidad Miguel Hernández, Av. Universidad s/n, E03202 Elche, Spain*
*[*]jcapmany@dcom.upv.es*

**Abstract:** We present a quantum model for electro-optic amplitude modulation, which is built upon quantum models of the main photonic components that constitute the modulator, that is, the guided-wave beamsplitter and the electro-optic phase modulator and accounts for all the different available modulator structures. General models are developed both for single and dual drive configurations and specific results are obtained for the most common configurations currently employed. Finally, the operation with two-photon input for the control of phase-modulated photons and the important topic of multicarrier modulation are also addressed.

## 1. Introduction

The electro-optic amplitude modulator (EOM) is the most popular modulating device employed in high speed optical communications systems featuring line rates in excess of 2.5 Gb/s [1,2]. It is also widely employed in analog photonic applications [3] and radio over fiber systems [4,5] where the modulating subcarriers are located in the RF, microwave and millimeter –wave regions of the electromagnetic spectrum.

The EOM is built by embedding one or two electro-optic phase modulators into a Mach-Zehnder interferometric setup closed by two guided-wave beamsplitters [5,6]. For instance, in Fig. 1 we show the most common EOM designs that are encountered in practice [5].

EOMs with only one internal phase modulator such as (B), (D), (F) in Fig. 1 are known as single drive or asymmetric modulators, while EOMs with two internal phase modulators, such as (A), (C) and (E) in Fig. 1 are known as dual drive modulators. For each phase modulator there are two ports, one for the modulator DC bias voltage and another to inject the time varying modulating signal. Designs (A) and (B) correspond to Y-Branch modulators, where both the input and output beamsplitters in the Mach-Zehnder interferometric structure are Y-Branch power splitters. These are the most common commercial devices. Designs (C) and (D) represent the so-called DC-Modulators where both the input and output beamsplitters in the Mach-Zehnder interferometric structure are guided wave directional couplers. These modulators bring the added value of an extra input and an extra output port but are more expensive to produce since the fabrication of a symmetric directional coupler is more challenging as compared to the Y-Branch power splitter which particularly simple to implement in integrated fashion [7,8]. DC modulators were common in the early days of integrated optics but today, although available upon request from different vendors they are not a mainstream commercial product. Finally, designs (E) and (F) correspond to hybrid Y-Branch DC modulators which feature one input and two outputs. These modulators, which can be found upon request in the market, are common for Cable TV applications [5], where the dual output permits a first signal splitting in the broadcasting header. The operation and design principles of the EOM under classical conditions are quite well established and understood and the interested reader can find useful information in innumerable references in the literature [3,9].

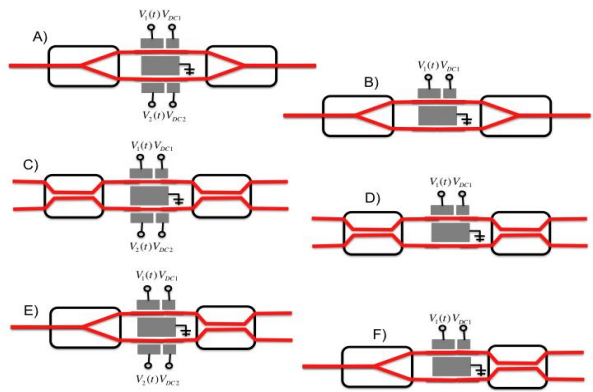

Fig. 1. Different possible layouts for amplitude electro-optic modulators. After [5].

As far as its operation under quantum regime there is however little if any work reported in the literature but nevertheless, the use of electro-optic modulation is finding increasing application in several areas including the control of quantum measurements [10], sources of approximated single-photon states in quantum key distribution (QKD) systems [11], frequency-coded [12,13] and subcarrier multiplexed [14] QKD systems or, more recently, for tailoring the wave function of heralded photons [15]. It is thus, an objective of this paper, to provide an accurate modeling of this device under such conditions. To do so, we have organized the paper as follows: In Section 2 we briefly present the existing quantum models for the building blocks of the EOM, that is, the beamsplitter and the phase modulator. In Section 3 we first develop the general transformation equations for the cases of single photon and coherent input states. The general results are then specialized in Section 4 as examples for several popular designs and operation modes of EOMs. In section 5 we briefly analyze in detail the operation of DC modulators under two-photon inputs, and show how they can be used to interpolate between two-port entangled and separable states composed of phase-modulated photons. In section 6 we develop the equation for the important case of multitone radiofrequency modulation. Finally the main summary and conclusions of the paper are presented in Section 6.

## 2. Quantum models for the beamsplitter, directional coupler, Y-branch power splitter and electro-optic phase modulator

### 2.1 Bulk-optics beamsplitter

The operation of the bulk optics BS under quantum regime can be found in several texts and references in the literature [10,16–19]. For instance, in [16] and [10,17,18] an excellent description based on the Heisenberg picture is developed. Here we focus as well on a description based on the Schrödinger picture that, for reasons to be apparent later in this chapter, is more convenient to our purposes of describing the operation of more complex photonic devices such as the amplitude electro-optic modulator. We then specialize the description to the two more prominent guided-wave versions of the device [6–8]; the Directional Coupler (DC) and the Y-Branch power splitter (YB) which are employed in the EOM. To describe the quantum operation of the optical beamsplitter we will refer to Fig. 2. The upper part of the figure describes the general framework of the Hilbert spaces and states that characterize both the input and the output of the BS. We will assume that the input is given by the product state $|in\rangle = |\Psi_1\rangle_1 \otimes |\Psi_2\rangle_2$, where $|\Psi_j\rangle_j$ belongs to $H_j$, the multimode Hilbert space corresponding to states of input port $j = 1,2$ (Although the Hilbert space is assumed multimode, the BS is a singlemode device and does not mix frequencies. So it is enough to analyze its operation over a single mode. Furthermore, we assume polarization independent operation.)

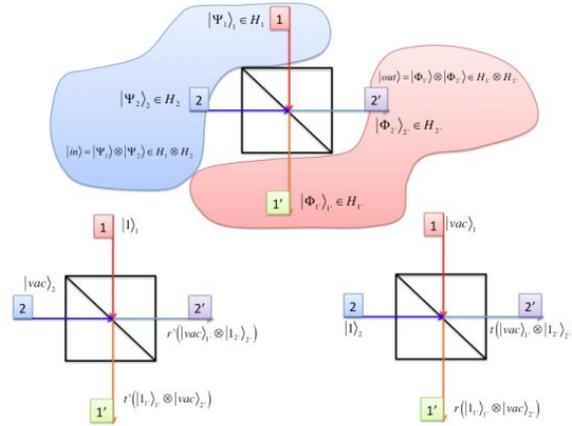

Fig. 2. Quantum state labeling (upper) and single photon behavior (lower) of an optical beamsplitter.

In the same way, an output state from the BS is given by $|out\rangle = |\Phi_1\rangle_{1'} \otimes |\Phi_2\rangle_{2'}$, with $|\Phi_{j'}\rangle_{j'} \in H_{j'}$ the Hilbert space corresponding to the states of output port $j' = 1',2'$. Upon these definitions, the action of the BS in the Heisenberg picture is described by an unitary scattering operator $\hat{S}_{BS}: |in\rangle \to |out\rangle$ which leaves invariant the vacuum state, $\hat{S}_{BS}|vac\rangle = |vac\rangle$ and such that, with the usual notations for the creation operators $\hat{a}_1^\dagger = \hat{a}^\dagger \otimes \mathbf{1}$ and $\hat{a}_2^\dagger = \mathbf{1} \otimes \hat{a}^\dagger$, its action can be expressed in matrix form:

$$\hat{S}_{BS}\begin{pmatrix} \hat{a}_1^\dagger \\ \hat{a}_2^\dagger \end{pmatrix}\hat{S}_{BS}^\dagger = \begin{pmatrix} t' & r' \\ r & t \end{pmatrix} \cdot \begin{pmatrix} \hat{a}_1^\dagger \\ \hat{a}_2^\dagger \end{pmatrix}, \quad (1)$$

where $t'$ and $r'$ define the field transmission and reflection coefficients for a classical signal fed through port 1 and correspondingly $t$ and $r$ are those for inputs from port 2. The explicit form for the bulk-optics beamsplitter scattering operator is [16]:

$$\hat{S}_{BS} = \exp(j\theta \hat{J}_1) = \exp[j\theta(\hat{a}_1^\dagger \hat{a}_2 + \hat{a}_2^\dagger \hat{a}_1)/2], \quad (2)$$

where $\theta$ determines the power division fraction between the output ports of the device ($\theta = \pi/2$ for a 3-dB beamsplitter), and definition Eq. (10) is consistent with Eq. (1) provided that:

$$t = t' = \cos(\theta/2) \quad r = r' = j\sin(\theta/2). \quad (3)$$

The action over single-mode displacement operators is given by [16]

$$\hat{S}_{BS}\left(\hat{D}(\alpha) \otimes \mathbf{1}\right)\hat{S}_{BS}^\dagger = \hat{D}(\alpha t') \otimes \hat{D}(\alpha r')$$
$$\hat{S}_{BS}\left(\mathbf{1} \otimes \hat{D}(\alpha)\right)\hat{S}_{BS}^\dagger = \hat{D}(\alpha r) \otimes \hat{D}(\alpha t), \quad (4)$$

from which we have the general transformation rule for arbitrary single-mode coherent-state inputs in both ports:

$$\hat{S}_{BS}\left(|\alpha\rangle_1 \otimes |\beta\rangle_2\right) = |\alpha t' + \beta r\rangle_{1'} \otimes |\alpha r' + \beta t\rangle_{2'}, \quad (5)$$

where $|\alpha\rangle = \hat{D}(\alpha)|vac\rangle$.

## 2.2 Directional coupler

The most widespread version of the optical beamsplitter in guided-wave format is the directional or 2 × 2 coupler [6–9]. It is composed on two input and two output optical fibers or integrated input waveguides. Signal coupling is achieved on the central part of the device by creating the adequate conditions such that the evanescent fields of the fundamental guided mode in one waveguide can excite the fundamental guided mode in the other and vice-versa. Different techniques to achieve signal coupling and for analyzing the device performance and carry out a proper design have been developed in the last 20 years and are quite well known and understood [6]. The beamsplitting power ratio of the 2 × 2 coupler is characterized by its coupling constant $k$ [6] that defines the fraction of power coupling (i.e crossing) from one waveguide to the other. Another characteristic is the $\pi/2$ phase shift that the optical field experiences when coupling from one waveguide to the other, which means that the reflection coefficients $r$, $r'$ in Eq. (3) are imaginary. The results for the general bulk-optics beamsplitter can thus be employed to describe the 2 × 2 directional coupler by making:

$$t = t' = \sqrt{1-k} \qquad r = r' = j\sqrt{k}. \tag{6}$$

## 2.3 Y-branch power splitter

The Y-Branch or 1 × 2 Power Splitter is another popular guided-wave implementation of the optical beamsplitter although it is more commonly employed in its integrated optics version than in optical fiber format [8]. As the 2 × 2 directional coupler it is characterized by its coupling constant $k$ that again, defines the fraction of power coupling (i.e crossing) from one waveguide to the other. However, in this case, there is no phase shift experienced by the optical field when coupling from one waveguide to the other. The behavior of the Y-Branch splitter under classical regime is shown in Fig. 3.

One distinctive feature of the Y-Branch is that it only has one physical input port. To characterize the device as a beamsplitter [18], we need to assume that there is a second input port although with no physical access from the classical point of view (on the reverse operation this port corresponds to the radiated field due to the antisymmetric supermode [8]). However, it should be borne in mind that under quantum regime this port is accessible by a vacuum state [16].

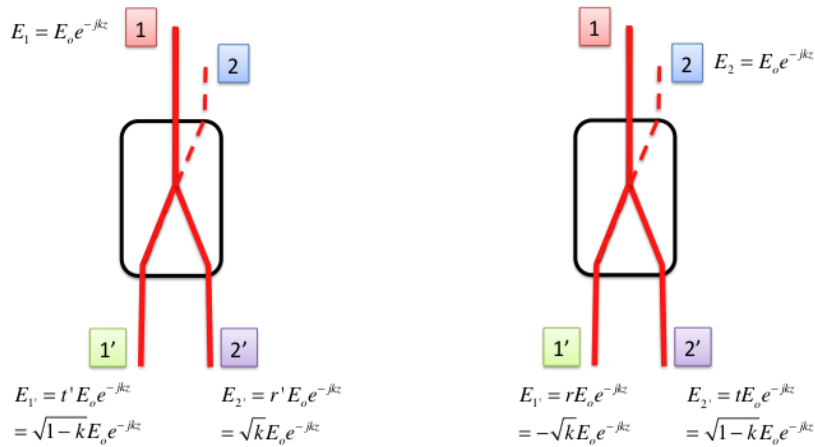

Fig. 3. Classical behavior of Y-branch guided-wave power splitter.

The energy conservation and reciprocity relationships can be summarized as:

$$|t'|^2 + |r'|^2 = |t|^2 + |r|^2 = 1 \qquad r^*t' + r't^* = 0. \qquad (7)$$

Now, if $t, t', r$ and $r'$ are real and $k$ characterizes the fraction of power coupling (i.e crossing) from one waveguide to the other then the above relationships are verified provided that:

$$t = t' = \sqrt{1-k} \qquad r' = \sqrt{k} = -r, \qquad (8)$$

so:

$$\hat{S}_{YB} \begin{pmatrix} \hat{a}_1^\dagger \\ \hat{a}_2^\dagger \end{pmatrix} \hat{S}_{YB}^\dagger = \begin{pmatrix} \sqrt{1-k} & \sqrt{k} \\ -\sqrt{k} & \sqrt{1-k} \end{pmatrix} \cdot \begin{pmatrix} \hat{a}_1^\dagger \\ \hat{a}_2^\dagger \end{pmatrix}. \qquad (9)$$

Alternatively, one can employ an explicit form for the Y-Branch power splitter scattering operator, which in this case is given by a different expression than that of Eq. (2):

$$\hat{S}_{YB} = \exp(-j\theta \hat{J}_2) = \exp[-\theta(\hat{a}_1^\dagger \hat{a}_2 - \hat{a}_2^\dagger \hat{a}_1)/2]. \qquad (10)$$

Again, it can be checked that Eq. (10) verifies the transformation given by Eq. (9) if:

$$t = t' = \sqrt{1-k} = \cos(\theta/2) \qquad r = -r' = -\sqrt{k} = -\sin(\theta/2). \qquad (11)$$

*2.4 Electro-optic phase modulator*

Quantum models for the electro-optic phase modulator have been derived in detail in [20,21]. Figure 4 shows a typical layout of the device. It consists of a dielectric waveguide and two electrodes placed at both sides. The dielectric material is subject to the electro-optic effect and the voltage applied to the electrodes changes linearly the refractive index undergone by one of the two possible input linear polarizations while not affecting the refractive index in the other linear polarization.

Using the overcomplete basis of coherent states and by analogy with the classical behavior it can be shown [21] that the modulator action subject to sinusoidal radiofrequency modulation is described by a unitary operator $\hat{S}_{PM}$ that couples modes with different frequencies. Such modulating tone at frequency $\Omega$ is given by $V(t) = V_{DC} + V_m\cos(\Omega t + \theta)$.where $\varphi_b = \pi V_{DC}/V_\pi$ is a static phase shift related to the modulator bias voltage normalized by the modulator $V_\pi$ parameter, $m = \pi V_m/V_\pi$ represents the radiofrequency modulation index. The optical output shows frequencies $\omega_0 + k\Omega$, with $k$ the integer determining the order of the output optical sideband. After quantizing the travelling-wave radiation in a length $L$, allowed frequencies become multiples of the quantity $2\pi v/L$, where $v$ is the speed of light in the medium. This implies that the incoming optical radiation can be described by index $n_0 > 0$ such that $\omega_0 = 2\pi n_0 v/L$, and correspondingly the radio-frequency $\Omega$ is determined by integer $N > 0$ with $\Omega = 2\pi Nc/L$. The scattering between waves with positive frequency is then given by:

$$\hat{S}_{PM} \hat{a}_{n_0}^\dagger \hat{S}_{PM}^\dagger = \sum_{q=1}^\infty C_q(q_0) \hat{a}_{qN-r_0}^\dagger. \qquad (12)$$

Here the coefficients

$$C_q(q_0) = e^{j\varphi_b}(je^{j\theta})^{q-q_0} \left[ J_{q-q_0}(m) - (-1)^{q_0} J_{q+q_0}(m) \right] \cong e^{j\varphi_b}(je^{j\theta})^{q-q_0} J_{q-q_0}(m), \qquad (13)$$

represent the scattering amplitudes from the input frequency represented by a mode number $n_0 = q_0 N - r_0$ ($q_0$ and $r_0$ being integers with $0 \leq r_0 < N$) to an output mode of frequency $qN-r_0$. In the last part of Eq. (13) we have introduced the so-called optical limit which recover the usual expressions of the classical theory of electro-optic phase modulation: when the incoming radiation is optical ($n_0 \gg 0$) the second Bessel function in Eq. (13) can be neglected and Eq. (12) can be written as:

$$\hat{S}_{PM} \hat{a}^{\dagger}_{n_0} \hat{S}^{\dagger}_{PM} \cong e^{j\varphi_b} \sum_{s=1-q_o}^{\infty} (je^{j\theta})^s J_s(m) \hat{a}^{\dagger}_{n_0+sN}, \qquad (14)$$

and then $C_q(q_0)$ represents the amplitude of the scattering from the incoming frequency to its sideband of order $q-q_0$.

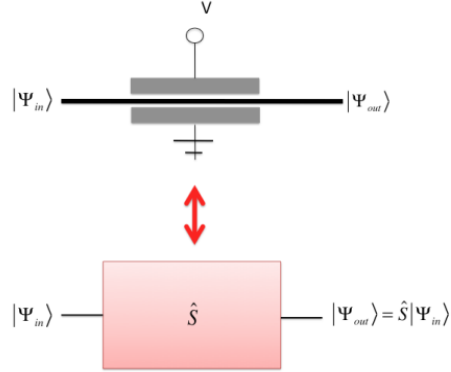

Fig. 4. Typical configuration of a waveguide electro-optic phase modulator. (Lower) Black-box representation of the phase modulator under quantum regime.

The phase modulator can be described by [21]:

$$\hat{S}_{PM} = \exp\left[ j(\chi \hat{T}_N + \chi^* \hat{T}^{\dagger}_N + \varphi_b \hat{N}_{ph}) \right], \qquad (15)$$

where $\chi = e^{j\theta} m/2$ and

$$\hat{T}_N = \sum_{m=1}^{\infty} \hat{a}^{\dagger}_{m+N} \hat{a}_m \qquad \hat{N}_{ph} = \hat{T}_0 = \sum_{m=1}^{\infty} \hat{a}^{\dagger}_m \hat{a}_m, \qquad (16)$$

### 3. General quantum model for the electro-optic amplitude modulator

We consider a general structure of the amplitude modulator as shown in Fig. 5. The layout is obviously an abstract representation as the beamsplitters are of the guided wave type [5,6,8]. However, the representation of Fig. 5 clearly shows the three building blocks of the modulator; an input beamsplitter opening the two paths of the interferometer, two different paths (labeled 1 and 2 in the figure) which include, each one, a phase modulator and an output beamsplitter which closes the interferometer and provides two possible output ports. Now, the operator describing the action of the EOM, $\hat{S}_{EOM}$, is given by:

$$\hat{S}_{EOM} = \hat{S}_{BS_o} \hat{S}_{PM} \hat{S}_{BS_i} = \hat{S}_{BS_o} (\hat{S}_{PM_1} \otimes \hat{S}_{PM_2}) \hat{S}_{BS_i}, \qquad (17)$$

where $\hat{S}_{PM1}$ and $\hat{S}_{PM2}$ represent the scattering operators corresponding, respectively, to the phase modulators located in the lower and upper paths of the layout of Fig. 5. $\hat{S}_{EOM}$ is unitary, since it is the product of three unitary operators, so that $\hat{S}_{EOM} \hat{S}^{\dagger}_{EOM} = \hat{S}^{\dagger}_{EOM} \hat{S}_{EOM} = \mathbf{1} \otimes \mathbf{1}$. We also mention that, although we use this complete layout to develop the model for the amplitude modulator, the equations developed here can and will be particularized to the case of asymmetric modulators as well where, say, $\hat{S}_{PM1}$ is substituted by the identity operator.

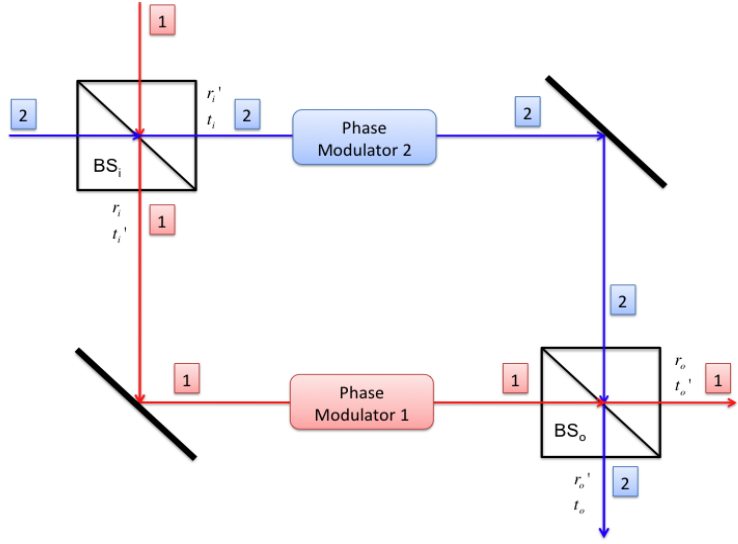

Fig. 5. Generic layout of an electro-optic amplitude modulator.

*3.1 Single Photon input states*

We shall start with the characterization of the EOM response to single photon inputs, that is, we wish to evaluate:

$$\hat{S}_{EOM}\left(|1\rangle_n \otimes |vac\rangle\right) = \hat{S}_{EOM}\left(\hat{a}_n^\dagger \otimes \mathbf{1}\right)|vac\rangle \otimes |vac\rangle = \hat{S}_{EOM}\left(\hat{a}_n^\dagger \otimes \mathbf{1}\right)\hat{S}_{EOM}^\dagger |vac\rangle \otimes |vac\rangle$$
$$\hat{S}_{EOM}\left(|vac\rangle \otimes |1\rangle_n\right) = \hat{S}_{EOM}\left(\mathbf{1} \otimes \hat{a}_n^\dagger\right)|vac\rangle \otimes |vac\rangle = \hat{S}_{EOM}\left(\mathbf{1} \otimes \hat{a}_n^\dagger\right)\hat{S}_{EOM}^\dagger |vac\rangle \otimes |vac\rangle,$$
(18)

where we have used that, since all the operators in (17) leave invariant the vacuum state, $\hat{S}_{EOM}$ also leaves it invariant. Using the results derived in section 2 we obtain, for instance:

$$\hat{S}_{EOM}\left(\hat{a}_n^\dagger \otimes \mathbf{1}\right)\hat{S}_{EOM}^\dagger = t_i' t_o' \left(\hat{S}_{PM_1}\hat{a}_n^\dagger \hat{S}_{PM_1}^\dagger \otimes \mathbf{1}\right) + r_i' r_o \left(\hat{S}_{PM_2}\hat{a}_n^\dagger \hat{S}_{PM_2}^\dagger \otimes \mathbf{1}\right)$$
$$+ t_i' r_o' \left(\mathbf{1} \otimes \hat{S}_{PM_1}\hat{a}_n^\dagger \hat{S}_{PM_1}^\dagger\right) + r_i' t_o \left(\mathbf{1} \otimes \hat{S}_{PM_2}\hat{a}_n^\dagger \hat{S}_{PM_2}^\dagger\right),$$
(19)

a transformation that possesses a straightforward physical interpretation. The first term in the right member represents the transformation corresponding to a state entering the EOM through port 1, transmitted by the input beamsplitter $BS_i$ to the lower path, being modulated by $PM_1$ and finally transmitted by the output beamsplitter $BS_o$ to output port 1. The second term represents the transformation corresponding to a state entering the EOM through port 1, transmitted by the input beamsplitter $BS_i$ to the lower path, being modulated by $PM_1$ and finally reflected by the output beamsplitter $BS_o$ to output port 2. The third term represents the transformation corresponding to a state entering the EOM through port 1, reflected by the input beamsplitter $BS_i$ to the upper path, being modulated by $PM_2$ and finally reflected by beamsplitter $BS_o$ to output port 1. The fourth and last term represents the transformation corresponding to a state entering the EOM through port 1, reflected by the input beamsplitter $BS_i$ to the upper paths, being modulated by $PM_2$ and finally transmitted by the output beamsplitter $BS_o$ to output port 2. Similarly we obtain:

$$\hat{S}_{EOM}\left(\mathbf{1}\otimes\hat{a}_n^\dagger\right)\hat{S}_{EOM}^\dagger = r_i't_o'\left(\hat{S}_{PM_1}\hat{a}_n^\dagger\hat{S}_{PM_1}^\dagger\otimes\mathbf{1}\right)+t_ir_o\left(\hat{S}_{PM_2}\hat{a}_n^\dagger\hat{S}_{PM_2}^\dagger\otimes\mathbf{1}\right)$$
$$+r_ir_o'\left(\mathbf{1}\otimes\hat{S}_{PM_1}\hat{a}_n^\dagger\hat{S}_{PM_1}^\dagger\right)+t_it_o\left(\mathbf{1}\otimes\hat{S}_{PM_2}\hat{a}_n^\dagger\hat{S}_{PM_2}^\dagger\right), \quad (20)$$

with a similar physical interpretation for its four terms but taking into account that now the input is in port 2. Although Eqs. (19) and (20) have been derived for a perfectly balanced interferometer they hold for unbalanced structures since any phase imbalance between the upper and the lower arms can be incorporated into the DC bias term $\varphi_b$.

To proceed further we do compact the notation by naming [see Eqs. (12) and (13)]:

$$\hat{b}_{n_0}^\dagger \equiv \hat{S}_{PM_1}\hat{a}_{n_0}^\dagger\hat{S}_{PM_1}^\dagger = \sum_{q=1}^\infty C_q(q_0)\hat{a}_{qN_1-r_1}^\dagger \qquad \hat{c}_{n_0}^\dagger \equiv \hat{S}_{PM_2}\hat{a}_{n_0}^\dagger\hat{S}_{PM_2}^\dagger = \sum_{q=1}^\infty \bar{C}_q(\bar{q}_0)\hat{a}_{qN_2-r_2}^\dagger, \quad (21)$$

where $\hat{b}^\dagger$ and $\hat{c}^\dagger$ have the interpretation of creators of phase-modulated photons. Referring to Eq. (20) we will assume in general that the frequencies of the RF signals modulating each of the two phase modulators are different (given by mode indexes $N_1$ and $N_2$, respectively). Thus, following the same discussion as after Eqs. (12) and (13) we have $n_0 = q_0 N_1 - r_1$ for phase modulator 1 and $n_0 = \bar{q}_0 N_2 - r_2$ for phase modulator 2. Furthermore, both modulators might show different characteristics (for example, they might be biased at different voltages), so we employ $C_q(x)$ and $\bar{C}_q(x)$ to express the fact that the coefficients given by Eq. (13) might be different for both modulators even if they are modulated by the same RF frequency.

Use of Eq. (21) in Eqs. (19) and (20) yields the expression of the transition amplitudes for a single-photon input to each of the multimode output in each output port as a function of BS modulator couplings. Although general, those expressions are rather involved and of limited use in practice. Simpler expressions result in the especially representative case of the same microwave frequency of the modulating signal in both RF ports so $N_1 = N_2 = N$, $r_1 = r_2 = r$, $q_0 = \bar{q}_0$. In this case Eqs. (19) and (20) reduce to:

$$\hat{S}_{EOM}\left(\hat{a}_n^\dagger\otimes\mathbf{1}\right)\hat{S}_{EOM}^\dagger = \left(\sum_{q=1}^\infty\left[t_it_o'C_q(q_0)+r_ir_o'\bar{C}_q(q_0)\right]\hat{a}_{qN-r}^\dagger\right)\otimes\mathbf{1}$$
$$+\mathbf{1}\otimes\left(\sum_{q=1}^\infty\left[t_ir_o'C_q(q_0)+r_it_o\bar{C}_q(q_0)\right]\hat{a}_{qN-r}^\dagger\right), \quad (22)$$

$$\hat{S}_{EOM}\left(\mathbf{1}\otimes\hat{a}_n^\dagger\right)\hat{S}_{EOM}^\dagger = \left(\sum_{q=1}^\infty\left[r_it_o'C_q(q_0)+t_ir_o\bar{C}_q(q_0)\right]\hat{a}_{qN-r}^\dagger\right)\otimes\mathbf{1}$$
$$+\mathbf{1}\otimes\left(\sum_{q=1}^\infty\left[r_ir_o'C_q(q_0)+t_it_o\bar{C}_q(q_0)\right]\hat{a}_{qN-r}^\dagger\right). \quad (23)$$

The interpretation of these formulas is now straightforward: due to the action of the phase modulators, an input photon couples to the same ladder of sideband modes $\hat{a}_{qN-r}^\dagger$ in each output port. The particular value of the transition amplitude depends on an interference term, i.e., on the coherent sum of the transition amplitudes $C_q$ and $\bar{C}_q$ set by each phase modulator, which can be tuned independently by different configuration of bias ($\varphi_{b1}$, $\varphi_{b2}$), radio-frequency phase ($\theta_1$, $\theta_2$) and modulation indices ($m_1$, $m_2$). This interference term determining the coupling to the output optical sidebands is finally weighted (in amplitude and phase) by the corresponding beamsplitter's reflection and transmission coefficients.

### 3.2 Coherent input states

In this case we need to evaluate:

$$\hat{S}_{EOM}(|\alpha_{n_0}\rangle_{n_0} \otimes |vac\rangle) = \hat{S}_{EOM}(\hat{D}_{n_0}(\alpha_{n_0}) \otimes \mathbf{1})\hat{S}^\dagger_{EOM}(|vac\rangle \otimes |vac\rangle)$$
$$\hat{S}_{EOM}(|vac\rangle \otimes |\alpha_{n_0}\rangle_{n_0}) = \hat{S}_{EOM}(\mathbf{1} \otimes \hat{D}_{n_o}(\alpha_{n_o}))\hat{S}^\dagger_{EOM}(|vac\rangle \otimes |vac\rangle), \quad (24)$$

where $|\alpha_{n_o}\rangle_{n_o}$ represents the coherent state for mode number $n_0$. We are thus interested in computing the transformations. Following a similar procedure to that of subsection 3.1 we arrive after a straightforward but lengthy process to:

$$\hat{S}_{EOM}(\hat{D}_{n_0}(\alpha_{n_0}) \otimes \mathbf{1})\hat{S}^\dagger_{EOM} =$$
$$= \left[\bigotimes_{q=1}^{\infty} \hat{D}_{qN_1-r_1}\left(\alpha_{n_0} t_i' t_o C_q(q_0)\right) \hat{D}_{qN_2-r_2}\left(\alpha_{n_o} r_i' r_o \bar{C}_q(\bar{q}_0)\right)\right] \otimes \left[\bigotimes_{q=1}^{\infty} \hat{D}_{qN_1-r_1}\left(\alpha_{n_o} t_i' r_o' C_q(q_o)\right) \hat{D}_{qN_2-r_2}\left(\alpha_{n_o} r_i' t_o \bar{C}_q(\bar{q}_0)\right)\right], \quad (25)$$

where the brackets separate the multimode outputs in each of the two ports, and, similarly:

$$\hat{S}_{EOM}(\mathbf{1} \otimes \hat{D}_{n_0}(\alpha_{n_0}))\hat{S}^\dagger_{EOM}$$
$$= \left[\bigotimes_{q=1}^{\infty} \hat{D}_{qN_1-r_1}\left(\alpha_{n_o} r_i' t_o C_q(q_0)\right) \cdot \hat{D}_{qN_2-r_2}\left(\alpha_{n_o} t_i r_o \bar{C}_q(\bar{q}_0)\right)\right] \otimes \left[\bigotimes_{q=1}^{\infty} \hat{D}_{qN_1-r_1}\left(\alpha_{n_o} r_i' r_o' C_q(q_0)\right) \hat{D}_{qN_2-r_2}\left(\alpha_{n_o} t_i t_o \bar{C}_q(\bar{q}_0)\right)\right]. \quad (26)$$

Again, when the microwave frequency of the modulating signal is the same in both RF ports ($N_1 = N_2 = N$, $r_1 = r_2 = r$, $q_0 = \bar{q}_0$), the former equations reduce to:

$$\hat{S}_{EOM}(\hat{D}_{n_0}(\alpha_{n_0}) \otimes \mathbf{1})\hat{S}^\dagger_{EOM} =$$
$$= \left[\bigotimes_{q=1}^{\infty} \hat{D}_{qN-r}\left[\alpha_{n_0}\left(t_i' t_o C_q(q_0) + r_i' r_o \bar{C}_q(q_0)\right)\right]\right] \otimes \left[\bigotimes_{q=1}^{\infty} \hat{D}_{qN-r}\left[\alpha_{n_o}\left(t_i r_o' C_q(q_0) + r_i' t_o \bar{C}_q(q_0)\right)\right]\right], \quad (27)$$

$$\hat{S}_{EOM}(\mathbf{1} \otimes \hat{D}_{n_0}(\alpha_{n_0}))\hat{S}^\dagger_{EOM} =$$
$$= \left[\bigotimes_{q=1}^{\infty} \hat{D}_{qN-r}\left[\alpha_{n_0}\left(r_i' t_o C_q(q_0) + t_i r_o \bar{C}_q(q_0)\right)\right]\right] \otimes \left[\bigotimes_{q=1}^{\infty} \hat{D}_{qN-r}\left[\alpha_{n_o}\left(r_i r_o' C_q(q_0) + t_i t_o \bar{C}_q(q_0)\right)\right]\right]. \quad (28)$$

Comparison of Eqs. (27) and (28) with Eqs. (22) and (23) shows that the multimode coherent-state outputs are governed by the same interference term or coherent sum of phase-modulator settings (through $C_q$ and $\bar{C}_q$) weighted by the beamsplitter reflection and transmission coefficients. The situation is thus similar to that of a single beamsplitter or a passive interferometer [16], where the classical output amplitudes described by the product states Eqs. (27) and (28) are in direct correspondence of the transition amplitudes of the genuine quantum output state Eqs. (22) and (23) when the modulator is operated at single-photon level. In classical, coherent-state operation, the interference between modes is designed to produce outputs with specific temporal features, such as pulse trains with different duty cycle for digital communications or amplitude waveforms with distortion-free, linear response for analog communications. At single-photon level, the action of the modulator is best viewed in the spectral domain, where the device operates as an active, multimode beamsplitter, with different output transition amplitudes that can be tuned by each of the phase-modulator settings.

## 4. Particular results for selected EOM configurations

Despite the generality of the expressions developed in the previous section, in practical terms however the range of amplitude modulators which are available either on the shelf or upon request from manufacturers is restricted to a few number of designs, most of which have been shown in Fig. 1. We now proceed to specialize the results obtained in 3.1 and 3.2 to those configurations that are more commonly encountered in practice [5], restricting our analysis to single-photon operation.

*4.1 Y-Branch modulator with two modulating inputs*

The layout of this modulator is shown in Fig. 1(A). In practical devices the design is such that:

$$t_o = t_o^{'} = t_i = t_i^{'} = 1/\sqrt{2} \qquad r_i^{'} = r_o = -r_i = -r_o^{'} = 1/\sqrt{2}. \tag{29}$$

Restricting the analysis to the case when the microwave frequency of the modulating signal is the same in both RF ports, and a single-photon input at the only accessible port, we get

$$\hat{S}_{EOM}\left(\hat{a}_{n_o}^\dagger \otimes \mathbf{1}\right)\hat{S}_{EOM}^\dagger = \frac{1}{2}\left[\sum_{q=1}^{\infty}\left(C_q(q_0)+\bar{C}_q(q_0)\right)\hat{a}_{qN-r}^\dagger \otimes \mathbf{1}\right] + \frac{1}{2}\left[\mathbf{1} \otimes \sum_{q=1}^{\infty}\left(-C_q(q_0)+\bar{C}_q(q_0)\right)\hat{a}_{qN-r}^\dagger\right], \tag{30}$$

with, according to Eq. (13)

$$C_q(q_0) = e^{j\varphi_{b1}}(je^{j\theta_1})^{q-q_0}\left[J_{q-q_0}(m_1)-(-1)^{q_0}J_{q+q_0}(m_1)\right]\varsigma$$
$$\bar{C}_q(q_0) = e^{j\varphi_{b2}}(je^{j\theta_2})^{q-q_0}\left[J_{q-q_0}(m_2)-(-1)^{q_0}J_{q+q_0}(m_2)\right]. \tag{31}$$

Then, the response to a single photon input is given by:

$$\hat{S}_{EOM}|1\rangle_{n_0} \otimes |vac\rangle = \frac{1}{2}\left[\sum_{q=1}^{\infty}\left(C_q(q_0)+\bar{C}_q(q_0)\right)|1\rangle_{qN-r} \otimes |vac\rangle\right] + \frac{1}{2}\left[|vac\rangle \otimes \sum_{q=1}^{\infty}\left(-C_q(q_0)+\bar{C}_q(q_0)\right)|1\rangle_{qN-r}\right]. \tag{32}$$

Now, we specialize this output for two standard settings of this modulator through the definition of bias, modulation indices and radio-frequency phase. The fist case is the so called *Double Sideband (DSB) operation in quadrature* [4,5] which, due to the low harmonic distortion represents one of the most standard setting in analog communications. It is characterized by $m_1 = m_2 = m$, $\theta_1 = 0$, $\theta_2 = \pi$, $\varphi_{b1} = -\varphi_{b2} = \pi/2$. Substituted into Eq. (31) yields:

$$C_q(q_0) = (j)^{q-q_0+1}\left[J_{q-q_0}(m)-(-1)^{q_0}J_{q+q_0}(m)\right]$$
$$\bar{C}_q(q_0) = (-j)^{q-q_0+1}\left[J_{q-q_0}(m)-(-1)^{q_0}J_{q+q_0}(m)\right] = C_q^*(q_0), \tag{33}$$

so Eq. (30) is now:

$$\hat{S}_{EOM}\left(\hat{a}_{n_0}^\dagger \otimes \mathbf{1}\right)\hat{S}_{EOM}^\dagger = \sum_{q=1}^{\infty}\text{Re}[C_q(q_0)]\,\hat{a}_{qN-r}^\dagger \otimes \mathbf{1} \;+\; \mathbf{1} \otimes \sum_{q=1}^{\infty} -j\,\text{Im}[C_q(q_0)]\,\hat{a}_{qN-r}^\dagger. \tag{34}$$

The first term in the right member of Eq. (34) corresponds to the output field from the modulator, while the second term identifies the radiated field. Turning our attention to the output field term, we can see that the real part of $C_q(q_0)$ is zero if $q-q_o$ is even. This means that no modes corresponding to even harmonics are expected at the modulator output. This is a well-known characteristic of this particular modulator design under classical operation.

The second interesting operation regime is the *Single Sideband (SSB) operation* [5] in which by properly dephasing by 90° one of the input RF modulating tones one of the RF sidebands (upper or lower) is eliminated at the output of the modulator. This operation mode is also of widespread use in analog communication due to its resilience to link's dispersion. In this case the values of the parameters are $m_1 = m_2 = m$, $\theta_1 = 0$, $\theta_2 = \pi/2$, $\varphi_{b1} = \pi/2$ and $\varphi_{b2} = 0$, substituted into Eq. (31) yields:

$$C_q(q_0) = (j)^{q-q_0+1} \left[ J_{q-q_0}(m) - (-1)^{q_0} J_{q+q_0}(m) \right]$$
$$\bar{C}_q(q_0) = (j)^{2q-2q_0} \left[ J_{q-q_0}(m) - (-1)^{q_0} J_{q+q_0}(m) \right]. \quad (35)$$

Note that for the special case when $q = q_0-1$, that is, for the lower RF sideband one gets:

$$\bar{C}_{q_0-1}(q_0) = -C_{q_0-1}(q_0), \quad (36)$$

and thus, as expected, the contribution to the lower RF sideband is cancelled in the first term of the right hand-side member of Eq. (32). It can be readily checked that the upper sideband is cancelled if $\theta_1 = 0$, $\theta_2 = -\pi/2$ while keeping unaltered the rest of the parameters.

*4.2 Y-Branch modulator with one modulating input*

The layout of this modulator is shown in Fig. 1(B), and its action can be derived from the results of the previous subsection after substituting the action of the second phase modulator PM2, which is not present, by the identity. Specifically:

$$\hat{S}_{EOM}(\hat{a}^\dagger_{n_0} \otimes \mathbf{1})\hat{S}^\dagger_{EOM} = \frac{1}{2}\left[(\hat{S}_{PM_1}\hat{a}^\dagger_{n_0}\hat{S}^\dagger_{PM_1} + \hat{a}^\dagger_{n_0}) \otimes \mathbf{1}\right] + \frac{1}{2}\left[\mathbf{1} \otimes (-\hat{S}_{PM_1}\hat{a}^\dagger_{n_0}\hat{S}^\dagger_{PM_1} + \hat{a}^\dagger_{n_0})\right] =$$
$$= \frac{1}{2}\left[\left(\sum_{q=1}^{\infty}(C_q(q_0)\hat{a}^\dagger_{qN_1-r_1}) + \hat{a}^\dagger_{n_0}\right) \otimes \mathbf{1}\right] + \frac{1}{2}\left[\mathbf{1} \otimes \left(\sum_{q=1}^{\infty}(-C_q(q_0)\hat{a}^\dagger_{qN_1-r_1}) + \hat{a}^\dagger_{n_0}\right)\right], \quad (37)$$

and the response to a single photon input is:

$$\hat{S}_{EOM}|1\rangle_{n_0} \otimes |vac\rangle = \frac{1}{2}\left[\left((C_{q_0}(q_0)+1)|1\rangle_{n_0} + \sum_{q=1,q\neq q_o}^{\infty} C_{q_0}(q_0)|1\rangle_{qN-r}\right) \otimes |vac\rangle\right] +$$
$$+ \frac{1}{2}\left[|vac\rangle \otimes \left((-C_{q_0}(q_0)+1)|1\rangle_{n_0} - \sum_{q=1,q\neq q_o}^{\infty} C_q(q_o)|1\rangle_{qN-r}\right)\right]. \quad (38)$$

Here, except for the input mode index $n_0$, the output is determined by the phase modulator PM1 whose multimode output is split in the output beamsplitter. For $n_0$, the output is governed by the interference between the unmodulated photon and the phase-modulated photon at mode $n_0$, whose transition amplitude is $C_{q_0}(q_0)$.

*4.3 Hybrid Y-Branch modulator with two modulating inputs*

The layout of this modulator is shown in Fig. 1(E), where:

$$t_o = t_o^{'} = t_i = t_i^{'} = r_i^{'} = -r_i = 1/\sqrt{2} \qquad r_o = r_o^{'} = j/\sqrt{2}. \tag{39}$$

For this configuration, again, there is only one physically accessible input port (port 1), and the former equation transforms to:

$$\hat{S}_{EOM}(\hat{a}_{n_0}^\dagger \otimes \mathbf{1})\hat{S}_{EOM}^\dagger =$$
$$= \frac{1}{2}\left[\sum_{q=1}^{\infty}\left(C_q(q_0) + j\bar{C}_q(q_0)\right)\hat{a}_{qN-r}^\dagger \otimes \mathbf{1}\right] + \frac{1}{2}\left[\mathbf{1} \otimes \sum_{q=1}^{\infty}\left(jC_q(q_0) + \bar{C}_q(q_0)\right)\hat{a}_{qN-r}^\dagger\right]. \tag{40}$$

Here, the interference terms produced in each sideband by the modulator represent the sum of the transition amplitudes with an additional relative phase $j$.

## 5. Two-photon input

In the previous section, the focus has been the multimode transition capabilities of one-photon states allowed by conventional designs of EOM. However, as interferometric devices with control of the relative phase, modulation index and modulating driving tone in each arm, EOMs permit compact implementations of standard, single-mode, quantum effects. In fact, if the modulation indices of the phase modulators are set to zero, they behave as phase shifters after the control of the bias voltage. Then, tasks based on bulk optics interferometers can be directly translated to the DC modulators of type 1(C) and 1(D), since the four ports of the interferometer are physically accessible and the mathematical descriptions of beamsplitter and directional coupler are the same. However, proper selection and operation of the EOM permit the integration in the same device of phase modulator and interferometer, thus allowing a direct extension of certain operations to multimode, phase-modulated fields.

In this section we show how DC-modulators can be used to switch and interpolate between two-port entangled and separable states composed of phase-modulated, multimode photons. The transformation corresponding to an operator representing two input photons (one at each input port) is given by:

$$\hat{S}_{EOM}\left(\hat{a}_{n_0}^\dagger \otimes \hat{a}_{n_0}^\dagger\right)\hat{S}_{EOM}^\dagger = \hat{S}_{EOM}\left(\hat{a}_{n_0}^\dagger \otimes \mathbf{1}\right)\hat{S}_{EOM}^\dagger \hat{S}_{EOM}\left(\mathbf{1} \otimes \hat{a}_{n_0}^\dagger\right)\hat{S}_{EOM}^\dagger =$$
$$= t_i^{'}t_o^{'}r_i t_o\left(\hat{b}_{n_0}^\dagger \hat{b}_{n_0}^\dagger \otimes \mathbf{1}\right) + 2t_i^{'}t_o^{'}r_i r_o^{'}\left(\hat{b}_{n_0}^\dagger \otimes \hat{b}_{n_0}^\dagger\right) + t_i^{'}r_o^{'}r_i r_o^{'}\left(\mathbf{1} \otimes \hat{b}_{n_0}^\dagger \hat{b}_{n_0}^\dagger\right) + r_i^{'}r_o^{'}t_i t_o\left(\hat{c}_{n_0}^\dagger \hat{c}_{n_0}^\dagger \otimes \mathbf{1}\right) +$$
$$+ 2r_i^{'}r_o^{'}t_i t_o\left(\hat{c}_{n_0}^\dagger \otimes \hat{c}_{n_0}^\dagger\right) + r_i^{'}t_o^{'}t_i t_o\left(\mathbf{1} \otimes \hat{c}_{n_0}^\dagger \hat{c}_{n_0}^\dagger\right) +$$
$$+ \left[t_i t_i^{'} + r_i r_i^{'}\right]\left\{r_o t_o^{'}\left(\hat{b}_{n_0}^\dagger \hat{c}_{n_0}^\dagger \otimes \mathbf{1}\right) + t_o t_o^{'}\left(\hat{b}_{n_0}^\dagger \otimes \hat{c}_{n_0}^\dagger\right) + r_o r_o^{'}\left(\hat{c}_{n_0}^\dagger \otimes \hat{b}_{n_0}^\dagger\right) + t_o r_o^{'}\left(\mathbf{1} \otimes \hat{b}_{n_0}^\dagger \hat{c}_{n_0}^\dagger\right)\right\}. \tag{41}$$

In practice the input beamsplitter to all modulators is always balanced, so that $|t_i| = |t_i'| = |r_i| = |r_i'| = 1/\sqrt{2}$ which, combined with the reciprocity relations, yields $t_i t_i' + r_i r_i' = 0$. This means that both photons circulate in the same modulator's arm due to the well-know effect [16]. Introducing in Eq. (41) the coefficients of the DC-modulators,

$$t_o = t_o^{'} = t_i = t_i^{'} = 1/\sqrt{2} \qquad r_i^{'} = r_i = r_o = r_o^{'} = j/\sqrt{2}, \tag{42}$$

we get:

$$\hat{S}_{EOM}\left(\hat{a}_{n_0}^\dagger \otimes \hat{a}_{n_0}^\dagger\right)\hat{S}_{EOM}^\dagger = \frac{i}{4}\left(\hat{b}_{n_0}^\dagger \hat{b}_{n_0}^\dagger \otimes \mathbf{1}\right) - \frac{1}{2}\left(\hat{b}_{n_0}^\dagger \otimes \hat{b}_{n_0}^\dagger\right) - \frac{i}{4}\left(\mathbf{1} \otimes \hat{b}_{n_0}^\dagger \hat{b}_{n_0}^\dagger\right)$$
$$- \frac{i}{4}\left(\hat{c}_{n_0}^\dagger \hat{c}_{n_0}^\dagger \otimes \mathbf{1}\right) - \frac{1}{2}\left(\hat{c}_{n_0}^\dagger \otimes \hat{c}_{n_0}^\dagger\right) + \frac{i}{4}\left(\mathbf{1} \otimes \hat{c}_{n_0}^\dagger \hat{c}_{n_0}^\dagger\right), \tag{43}$$

so that both output photons are modulated (scattered) by the same phase modulator. Furthermore, if we consider set equal modulation indices and radio-frequency driving tones in both phase modulators, we get $\hat{c}_{n_o}^\dagger = \exp(j\Delta\varphi_b)\hat{b}_{n_o}^\dagger$, where $\Delta\varphi_b$ represents the bias voltage difference between the upper and lower phase modulators, we obtain

$$\hat{S}_{EOM}\left(\hat{a}_{n_0}^\dagger \otimes \hat{a}_{n_0}^\dagger\right)\hat{S}_{EOM}^\dagger = \frac{e^{j\Delta\varphi_b}}{2}\sin(\Delta\varphi_b)[\hat{b}_{n_0}^\dagger\hat{b}_{n_0}^\dagger \otimes \mathbf{1} - \mathbf{1} \otimes \hat{b}_{n_0}^\dagger\hat{b}_{n_0}^\dagger] - e^{j\Delta\varphi_b}\cos(\Delta\varphi_b)\,\hat{b}_{n_0}^\dagger \otimes \hat{b}_{n_0}^\dagger. \quad (44)$$

Now, the structure of the output state can be controlled by $\Delta\varphi_b$. For instance, if $\Delta\varphi_b = 0$ then the output is a product state, whereas for $\Delta\varphi_b = \pi/2$ the output state is entangled, and at this point the overall modulator behaves in a similar fashion to a balanced beamsplitter acting on phase-modulated states.

## 6. Multitone amplitude modulation

We turn now our attention to the modeling of the EOM subject to multitone RF signal modulation as we previously considered in [21] for the electro-optic phase modulator. The interest in this analysis is justified by the fact that this is the regime under which several quantum experimental setups [10] and subcarrier multiplexed quantum key distribution systems [4,22], operate. We will assume, as in [21] linear and small signal modulation conditions, that is, $m \ll 1$.

In principle, the general expressions [Eqs. (17)–(20)] are valid but Eq. (20) has to be modified according to Eq. (49) of [21] to:

$$\hat{b}_{n_0}^\dagger \equiv \hat{S}_{PM_1}\hat{a}_{n_0}^\dagger\hat{S}_{PM_1}^\dagger = e^{j\varphi_{b1}}\hat{a}_{n_0}^\dagger + e^{j\varphi_{b1}}\sum_{k=1}^{M}m_{1k}\left[je^{j\theta_{1k}}\hat{a}_{n_0+N_{1k}}^\dagger + je^{-j\theta_{1k}}\hat{a}_{n_0-N_{1k}}^\dagger\right]$$

$$\hat{c}_{n_0}^\dagger \equiv \hat{S}_{PM_1}\hat{a}_{n_0}^\dagger\hat{S}_{PM_1}^\dagger = e^{j\varphi_{b2}}\hat{a}_{n_0}^\dagger + e^{j\varphi_{b2}}\sum_{k=1}^{M}m_{2k}\left[je^{j\theta_{2k}}\hat{a}_{n_0+N_{2k}}^\dagger + je^{-j\theta_{2k}}\hat{a}_{n_0-N_{2k}}^\dagger\right]. \quad (45)$$

Here $M$ stands for the number of RF tones in the modulating signal. For SCM-QKD systems in particular we are principally interested in the behavior under a coherent state input, since this is the state characterizing the output of a faint pulsed laser source. For this case we can follow the same process as in section 5 of [21] and sections 3 and 4 but taking into account the prescriptions given by Eq. (45) to find:

$$\hat{S}_{EOM}\left(\hat{D}_{n_0}\left(\alpha_{n_0}\right) \otimes \mathbf{1}\right)\hat{S}_{EOM}^\dagger = \hat{A} \otimes \hat{B}$$

$$\hat{A} = \hat{D}_{n_0}\left[\alpha_{n_0}\left(t_i't_o'e^{j\varphi_{b1}} + r_i'r_o'e^{j\varphi_{b2}}\right)\right]\bigotimes_{k=1}^{M}\hat{D}_{n_0\pm N_{1k}}\left[t_i't_o'm_{1k}j\alpha_{n_0}e^{j\varphi_{b1}}e^{\pm j\theta_{1k}}\right]\hat{D}_{n_0\pm N_{2k}}\left[r_i'r_o'm_{2k}j\alpha_{n_0}e^{j\varphi_{b2}}e^{\pm j\theta_{2k}}\right]$$

$$\hat{B} = \hat{D}_{n_0}\left[\alpha_{n_0}\left(t_i'r_o'e^{j\varphi_{b1}} + r_i't_o'e^{j\varphi_{b2}}\right)\right]\bigotimes_{k=1}^{N}\hat{D}_{n_0\pm N_{1k}}\left[t_i'r_o'm_{1k}j\alpha_{n_0}e^{j\varphi_{b1}}e^{\pm j\theta_{1k}}\right]\hat{D}_{n_0\pm N_{2k}}\left[r_i't_o'm_{2k}j\alpha_{n_0}e^{j\varphi_{b2}}e^{\pm j\theta_{2k}}\right], \quad (46)$$

and:

$$\hat{S}_{EOM}\left(\mathbf{1} \otimes \hat{D}_{n_0}\left(\alpha_{n_0}\right)\right)\hat{S}_{EOM}^\dagger = \hat{E} \otimes \hat{F}$$

$$\hat{E} = \hat{D}_{n_0}\left[\alpha_{n_0}\left(r_i't_o'e^{j\varphi_{b1}} + t_i'r_o'e^{j\varphi_{b2}}\right)\right]\bigotimes_{k=1}^{M}\hat{D}_{n_0\pm N_{1k}}\left[r_i't_o'm_{1k}j\alpha_{n_0}e^{j\varphi_{b1}}e^{\pm j\theta_{1k}}\right]\hat{D}_{n_0\pm N_{2k}}\left[t_i'r_o'm_{2k}j\alpha_{n_0}e^{j\varphi_{b2}}e^{\pm j\theta_{2k}}\right]$$

$$\hat{F} = \hat{D}_{n_0}\left[\alpha_{n_0}\left(r_i'r_o'e^{j\varphi_{b1}} + t_i't_o'e^{j\varphi_{b2}}\right)\right]\bigotimes_{k=1}^{N}\hat{D}_{n_0\pm N_{1k}}\left[r_i'r_o'm_{1k}j\alpha_{n_0}e^{j\varphi_{b1}}e^{\pm j\theta_{1k}}\right]\hat{D}_{n_0\pm N_{2k}}\left[t_i't_o'm_{2k}j\alpha_{n_0}e^{j\varphi_{b2}}e^{\pm j\theta_{2k}}\right]. \quad (47)$$

The former expressions can be simplified if the same microwave modulation frequencies are assumed for both RF ports. In this case:

$$\hat{A} = \hat{D}_{n_0}\left[\alpha_{n_0}\left(t_i't_o'e^{j\varphi_{b1}} + r_i'r_oe^{j\varphi_{b2}}\right)\right] \bigotimes_{k=1}^{M} \hat{D}_{n_0 \pm N_k}\left[j\alpha_{n_o}\left(t_i't_o'm_{1k}e^{j\varphi_{b1}}e^{\pm j\theta_{1k}} + r_i'r_om_{2k}e^{j\varphi_{b2}}e^{\pm j\theta_{2k}}\right)\right]$$

$$\hat{B} = \hat{D}_{n_0}\left[\alpha_{n_0}\left(t_i'r_o'e^{j\varphi_{b1}} + r_i't_oe^{j\varphi_{b2}}\right)\right] \bigotimes_{k=1}^{M} \hat{D}_{n_0 \pm N_k}\left[j\alpha_{n_o}\left(t_i'r_o'm_{1k}e^{j\varphi_{b1}}e^{\pm j\theta_{1k}} + r_i't_om_{2k}e^{j\varphi_{b2}}e^{\pm j\theta_{2k}}\right)\right]$$

$$\hat{E} = \hat{D}_{n_0}\left[\alpha_{n_0}\left(r_i't_o'e^{j\varphi_{b1}} + t_i'r_oe^{j\varphi_{b2}}\right)\right] \bigotimes_{k=1}^{M} \hat{D}_{n_0 \pm N_k}\left[j\alpha_{n_o}\left(r_i't_o'm_{1k}e^{j\varphi_{b1}}e^{\pm j\theta_{1k}} + t_i'r_om_{2k}e^{j\varphi_{b2}}e^{\pm j\theta_{2k}}\right)\right]$$

$$\hat{F} = \hat{D}_{n_0}\left[\alpha_{n_0}\left(r_i'r_o'e^{j\varphi_{b1}} + t_i't_oe^{j\varphi_{b2}}\right)\right] \bigotimes_{k=1}^{M} \hat{D}_{n_0 \pm N_k}\left[j\alpha_{n_o}\left(r_i'r_o'm_{1k}e^{j\varphi_{b1}}e^{\pm j\theta_{1k}} + t_i't_om_{2k}e^{j\varphi_{b2}}e^{\pm j\theta_{2k}}\right)\right].$$

(48)

These expressions are obviously general and can be particularized to the different modulator designs as we did in Section 4.

It is useful to compute the value of the mean fields corresponding to Eqs. (46)–(48). To do so we will assume that the input coherent state is fed to only one of the device ports (port 1) and assume equal modulation frequencies at both ports. In such a case, the output state is given by:

$$|\Psi\rangle = \hat{S}_{EOM}|\alpha_{n_0}\rangle_{n_0} \otimes |vac\rangle = \hat{S}_{EOM}(\hat{D}_{n_0}(\alpha_{n_o}) \otimes \mathbf{1})\hat{S}^{\dagger}_{EOM}|vac\rangle \otimes |vac\rangle = |\Psi_1\rangle_1 \otimes |\Psi_2\rangle_2$$

$$|\Psi_1\rangle_1 = \left|\alpha_{n_0}(t_i't_o'e^{j\varphi_{b1}} + r_i'r_oe^{j\varphi_{b2}})\right\rangle_{n_0} \bigotimes_{k=1}^{M} \left|j\alpha_{n_0}(t_i't_o'm_{1k}e^{j\varphi_{b1}}e^{\pm j\theta_{1k}} + r_i'r_om_{2k}e^{j\varphi_{b2}}e^{\pm j\theta_{2k}})\right\rangle_{n_o \pm N_k}$$

$$|\Psi_2\rangle_2 = \left|\alpha_{n_0}(t_i'r_o'e^{j\varphi_{b1}} + r_i't_oe^{j\varphi_{b2}})\right\rangle_{n_0} \bigotimes_{k=1}^{M} \left|j\alpha_{n_0}(t_i'r_o'm_{1k}e^{j\varphi_{b1}}e^{\pm j\theta_{1k}} + r_i't_om_{2k}e^{j\varphi_{b2}}e^{\pm j\theta_{2k}})\right\rangle_{n_o \pm N_k}.$$

(49)

To evaluate the mean field at the output port 1 of the modulator we consider the electric field operator given by [16]:

$$\hat{E}_1^+ = j\mathbf{x}\sum_{\omega}\xi(\omega)\hat{a}_{\omega}e^{-j\omega t} \otimes \mathbf{1} \qquad \xi(\omega) = \sqrt{\frac{\hbar\omega}{2\varepsilon_o V}},$$

(50)

where **x** is the unit vector relevant polarization. Then, applied to Eq. (51), yields an average field:

$$\langle\Psi|\hat{E}_1^+ + hc|\Psi\rangle = j\mathbf{x}\xi(\omega_0)\alpha_{n_0}\left(t_i't_o'e^{j\varphi_{b1}} + r_i'r_oe^{j\varphi_{b2}}\right)e^{-j\omega_0 t} +$$
$$+ j\mathbf{x}\sum_{k=1}^{M}\xi(\omega_0 \pm \Omega_k)j\alpha_{n_0}\left(t_i't_o'm_{1k}e^{j\varphi_{b1}}e^{\pm j\theta_{1k}} + r_i'r_om_{2k}e^{j\varphi_{b2}}e^{\pm j\theta_{2k}}\right)e^{-j(\omega_0 \pm \Omega_k)t} + cc.$$

(51)

Equation (51) shows the presence of photons in the mode corresponding to the input coherent state as well as in the sidebands corresponding to the *M* modulating RF subcarriers.

**6. Summary and conclusions**

We have presented a quantum model for electro-optic amplitude modulation, which is built upon quantum models of the main photonic components that constitute the EOM, that is, the guided-wave beamsplitter and the electro-optic phase modulator and accounts for all the different modulator structures which are currently available. General models have been developed both for single and dual drive configurations and specific results were obtained for the most common configurations currently employed. Finally, the operation with two-photon input for the control of phase-modulated photons and the important topic of multicarrier modulation are also addressed.

**Acknowledgements**

The authors wish to acknowledge the financial support of the Spanish Government through Project TEC2008-02606 and Project Quantum Optical Information Technology (QOIT), a CONSOLIDER-INGENIO 2010 Project; and also the Generalitat Valenciana through the PROMETEO research excellency award programme GVA PROMETEO 2008/092.